\documentclass[11pt]{article} 
\usepackage{amsmath}

\makeatletter
\@addtoreset{equation}{section}
\makeatother
\def\theequation{\thesection.\arabic{equation}}
\def\thefootnote{%
\fnsymbol{footnote}}

\newcommand{\dps}{\displaystyle }
\newcommand{\e }{\varepsilon }

\newcommand{\al }{\alpha }
\newcommand{\de }{\delta }

\newcommand{\ket }{\rangle }
\newcommand{\bra }{\langle }
\newcommand{\ga }{\gamma }
\newcommand{\la }{\lambda }
\newcommand{\Hw }{H_{\hbox{w}}} 
\newcommand{\im }{\hbox{Im}}
\newcommand{\re }{\hbox{Re}}
\newcommand{\La }{\Lambda}
\newcommand{\De }{\Delta}
\newcommand{\no }{\nonumber}
\newcommand{\kob }{\overline{K^0}}
\newcommand{\ko }{K^0}
\newcommand{\nub }{\overline{\nu}}

\newcommand{\nb }{\overline{n}}

\newcommand{\psib }{\overline{\psi}}
\newcommand{\Ab}{\overline{A}}
\begin{document}
\begin{flushright}
NUP-A-2002-3 \\
August 2002
\end{flushright}
~\\ ~\\ 
\begin{center}
\Large{$CP$, $T$ and $CPT$ Violations in the $\ko-\kob$ System \\ 
--- Present Status ---}
\end{center} 
~\\ ~\\ 
\begin{center}
Yoshihiro Takeuchi\footnote{E-mail address: yytake@phys.cst.nihon-u.ac.jp} and
S. Y. Tsai\footnote{E-mail address: tsai@phys.cst.nihon-u.ac.jp}
\end{center}

\begin{center}
{\it Institute of Quantum Science and Department of Physics \\ 
College of Science and Technology, Nihon University \\ 
Kanda-Surugadai, Chiyoda-ku, Tokyo 101-8308, Japan }
\end{center}
~\\ ~\\ 

\setcounter{footnote}{0}
\def\thefootnote{%
\alph{footnote}}
\vspace{1cm} 
\begin{abstract}
Possible violation of $CP$, $T$ and $CPT$ symmetries in the $\ko-\kob$ 
system is studied in a way as phenomenological and comprehensive as possible. For this 
purpose, we first introduce parameters which represent violation of these symmetries in 
mixing parameters and decay amplitudes in a convenient and well-defined way and, treating these parameters as small, derive formulas which relate them to the experimentally 
measured quantities. We then perform numerical analyses to derive constraints to these 
symmetry-violating parameters, with the latest data reported by KTeV Collaboration, NA48 
Collaboration and CPLEAR Collaboration, along with those compiled by Particle Data 
Group, used as inputs. The result obtained by CPLEAR Collaboration from an unconstrained 
fit to a time-dependent leptonic asymmetry, aided by the Bell-Steinberger relation, 
enables us to determine or constrain most of the parameters separately. It is shown among the other things that 
(1) $CP$ and $T$ symmetries are violated definitively at least at the level of $10^{-4}$ 
in $2\pi$ decays, 
(2) $CP$ and $T$ symmetries are violated at least at the level of $10^{-3}$ in the $\ko-\kob$ mixing, and 
(3) $CPT$ symmetry is at present tested to the level of $10^{-5}$ at the utmost. 
\end{abstract}

\newpage 

\section{Introduction}
Although, on the one hand, all experimental observations up to now are perfectly 
consistent with $CPT$ symmetry, and, on the other hand, the standard quantum field theory implies that this symmetry should hold exactly, continued experimental, phenomenological 
and theoretical studies of this and related symmetries are warranted. In this connection, we like to recall, on the one hand, that $CP$ symmetry is violated only at such a tiny level as $10^{-3}$ [1,2], while $CPT$ symmetry has been tested at best down to the level one order smaller [3] and, on the other hand, that some of the premises of the $CPT$ theorem, e.g., locality, are being challenged by, e.g., the superstring model.
 
In a series of papers [4-9], we studied possible violation of $CP$, $T$ and $CPT$ 
symmetries in the $\ko-\kob$ system from a phenomenological point of view. The procedure 
of our studies went as follows. We first introduced parameters which represented violation of $CP$, $T$ and $CPT$ symmetries in mixing and decay of $\ko$ and $\kob$ in a convenient and well-defined way 
and related them to the experimentally measured quantities. We then carried out 
numerical analyses, with and without the aid of the Bell-Steinberger relation [10] and 
with the available data on $2\pi$, $3\pi$, $\pi^+\pi^-\ga$ and $\pi\ell\nu_{\ell}$ decays used as inputs, to derive constraints to these symmetry-violating parameters. Along with 
the data compiled by Particle Data Group [11], the new results on $\re(\e'/\e)$ reported 
by KTeV Collaboration and NA48 Collaboration [12], as well as some of the results 
obtained by CPLEAR Collaboration [13-16], were taken into account.

The present work is an updated, revised and integrated version of the previous works, 
which is new particularly in the following points:

(1) In order to be as phenomenological and comprehensive as possible, along with the latest data reported by KTeV Collaboration [17] and by NA48 Collaboration [18] as well as those compiled by Particle Data Group [19], the result obtained by CPLEAR Collaboration from 
an unconstrained fit to a time-dependent leptonic asymmetry [20] is used as inputs in the numerical analyses.

(2) Relevant decay amplitudes are parametrized with phase ambiguities taken into account 
in an explicit way, and, in Appendix, issues related to rephasing and to phase conventions are discussed in somw detail.\footnote{See also Refs.[21,22].}

The paper is organized as follows. The theoretical framework used to describe 
the $\ko-\kob$ system [23] is recapitulated in Sect.2 and the experimentally measured quantities related to $CP$ violation in decay modes of interest to us are enumerated in 
Sect.3. We then, in Sect.4, parametrize the mixing parameters and decay amplitudes in a 
convenient and well-defined way and, on adopting a particular phase convention, give 
conditions imposed by $CP$, $T$ and/or $CPT$ symmetries on these parameters. In Sect.5, 
experimentally measured quantities are expressed in terms of the parameters defined, 
treating them as first order small. In Sect.6, with the relevant data used as inputs, 
numerical analyses are carried out to evaluate or constrain $CP$, $T$ and/or $CPT$ 
violating parameters separately as far as possible. The results of the analyses are 
summarized and some concluding remarks are given in  Sect.7. In Appendix, issues related to rephasing, rephasing-(non)invariance and phase conventions are discussed.

\section{The $\ko-\kob$ mixing and the Bell-Steinberger 
relation}
Let $|\ko\ket$ and $|\kob\ket$ be eigenstates of the strong interaction with strangeness $S=+1$ and $-1$, related to each other by $(CP)$, $(CPT)$ and $T$ operations as [4,21,22,24]
%
%
\begin{equation}
\left. 
	\begin{array}{cc}
	(CP)|\ko\ket = e^{i\al_K}|\kob\ket ~,& (CPT)|\ko\ket = e^{i\beta _K}|\kob\ket ~,\\ 
	(CP)|\kob\ket = e^{-i\al_K}|\ko\ket ~,& (CPT)|\kob\ket = e^{i\beta_K}|\ko\ket ~,\\ 
	T|\ko\ket = e^{i(\beta_K-\al_K)}|\ko\ket ~,&T|\kob \ket = e^{i(\beta_K+\al_K)}|\kob\ket ~.
	\end{array}
\right. 
\end{equation}
Note here that, given the first two where $\al_K$ and $\beta_K$ are arbitrary real parameters, the rest follow from the assumptions $(CP)T=T(CP)=(CPT)$, $(CP)^2=(CPT)^2=1$, and anti-linearity of $T$ and $(CPT)$.
When the weak interaction $\Hw$ is switched on, the $\ko$ and $\kob$ states decay into other states, generically denoted as $|n\ket$, and get mixed. The time evolution of the arbitrary state 
\[
|\Psi(t)\ket = c_1(t)|K_1\ket + c_2(t)|K_2\ket~,
\]
with
\[
|K_1\ket \equiv |\ko\ket~, \qquad |K_2\ket \equiv |\kob\ket~,
\]
is described by a Schr\"odinger-like equation [23,25] 
\[
i\frac{d}{dt}
|\Psi\ket = \La |\Psi\ket~,
\]
or
%
%
\begin{equation}
i\frac{d}{dt}
\left( 
	\begin{array}{c}
	c_1(t)\\ c_2(t) 
	\end{array}
\right) 
=\La 
\left( 
	\begin{array}{c}
	c_1(t)\\ c_2(t) 
	\end{array}
\right) ~.
\end{equation}
The operator or $2 \times 2$ matrix $\La$ may be written as
%
%
\begin{equation}
\La \equiv M - i\Gamma/2~,
\end{equation}
with $M$ (mass matrix) and $\Gamma$ (decay or width matrix) given, to the second 
order in $\Hw$, by
%
%
\begin{subequations}
\begin{eqnarray}
M_{ij} &\equiv& \bra K_i|M|K_j \ket \no \\
       &=& m_K\de_{ij} + \bra K_i|\Hw|K_j \ket \no \\ 
       & & \quad\qquad + \sum _n P \frac{\bra K_i|\Hw|n\ket \bra n|\Hw|K_j \ket}{m_K-E_n}~, \\
\Gamma_{ij} &\equiv& \bra K_i|\Gamma|K_j \ket \no \\
            &=& 2\pi \sum _n \bra K_i|\Hw|n\ket \bra n|\Hw|K_j \ket \de(m_K-E_n)~,\no \\
\end{eqnarray}
\end{subequations}
where the operator $P$ projects out the principal value. The two eigenstates of $\La$ and their respective eigenvalue may be written as 
%
%
\begin{subequations}
\begin{eqnarray}
|K_S\ket &=& \frac{1}{\sqrt{|p_S|^2+|q_S|^2}}\left( p_S|\ko \ket +q_S|\kob \ket \right) ~, \\
|K_L\ket &=& \frac{1}{\sqrt{|p_L|^2+|q_L|^2}}\left( p_L|\ko \ket -q_L|\kob \ket \right) ~; 
\end{eqnarray}
\end{subequations}
%
%
\begin{subequations}
\begin{eqnarray}
\la _S=m_S-i\frac{\ga _S}{2}~, \\
\la _L=m_L-i\frac{\ga _L}{2}~.
\end{eqnarray}
\end{subequations}
$m_{S, L}=\re (\la _{S, L})$ and $\ga _{S, L}=-2\im (\la _{S, L})$ are the mass
 and the total decay width of the $K_{S, L}$ state respectively. By definition,
 $\ga_S > \ga_L$ or $\tau_S < \tau_L$ ($\tau_{S,L} \equiv 1/\ga_{S,L}$), and 
the suffices $S$ and $L$ stand for "short-lived" and "long-lived" respectively. The eigenvalues $\la_{S,L}$ and the ratios of the mixing parameters 
$q_{S,L}/p_{S,L}$ are related to the elements of the mass-width matrix $\La$ as 
%
%
\begin{equation}
\la_{S,L} = \pm E + (\La_{11} + \La_{22})/2~,
\end{equation}
\begin{equation}
q_{S,L}/p_{S,L} = \La_{21}/[E \pm (\La_{11} - \La_{22})/2]~,
\end{equation}
where
%
%
\begin{equation}
E \equiv [\La_{12}\La_{21} + (\La_{11} - \La_{22})^2/4]^{1/2}.
\end{equation}

From the eigenvalue equation of $\La$, one may readily derive the well-known Bell-Steinberger relation [10]:
%
%
\begin{equation}
\left[ \frac{\ga_S+\ga_L}{2}-i(m_S-m_L) \right] \bra K_S|K_L\ket = \bra K_S|\Gamma |K_L\ket ~,
\end{equation}
where 
%
%
\begin{equation}
\bra K_S|\Gamma |K_L\ket = 2\pi \sum _n\bra K_S|\Hw|n\ket \bra n|\Hw |K_L\ket \de (m_K-E_n)~.
\end{equation}
%
\section{Decay modes}
\label{sec:3}
The $\ko$ and $\kob$ (or $K_S$ and $K_L$) states have many decay modes, among which we are interested in $2\pi$, $3\pi$, $\pi^+\pi^-\ga$ and semi-leptonic modes.
%
\subsection{$2\pi$ modes }
\label{sec:3.1}
The experimentally measured quantities related to $CP$ violation are $\eta _{+-}$ and $\eta _{00}$ defined by 
%
%
\begin{subequations}
\begin{equation}
\eta _{+-} \equiv |\eta _{+-}|e^{i\phi _{+-}} \equiv \frac{\bra \pi ^+\pi ^-,\mbox{outgoing}|\Hw |K_L\ket }{\bra \pi ^+\pi ^-,\mbox{outgoing}|\Hw |K_S\ket }~, 
\end{equation}
\begin{equation}
\eta _{00} \equiv |\eta _{00}|e^{i\phi _{00}} \equiv \frac{\bra \pi ^0\pi ^0,\mbox{outgoing}|\Hw |K_L\ket }{\bra \pi ^0\pi ^0,\mbox{outgoing}|\Hw |K_S\ket }~. 
\end{equation}
\end{subequations}
Defining
%
%
\begin{equation}
\omega \equiv \frac{\bra (2\pi )_2|\Hw |K_S\ket }{\bra (2\pi )_0|\Hw |K_S\ket }~,
\end{equation}
%
%
\begin{equation}
\eta _I \equiv |\eta_I|e^{i\phi_I} \equiv \frac{\bra (2\pi )_I|\Hw |K_L\ket }{\bra (2\pi )_I|\Hw |K_S\ket }~,
\end{equation}
where $I$=1 or 2 stands for the isospin of the $2\pi$ states, one gets
%
%
\begin{subequations}
\begin{eqnarray}
\eta _{+-} &=& \frac{\eta _0+\eta_2\omega '}{1+\omega '}~, \\
\eta _{00} &=& \frac{\eta _0-2\eta _2\omega '}{1-2\omega '}~,
\end{eqnarray}
\end{subequations}
where 
%
%
\begin{equation}
\omega ' \equiv \frac{1}{\sqrt{2}}\omega e^{i(\de _2-\de _0)}~,
\end{equation}
$\de _I$ being the S-wave $\pi \pi $ scattering phase shift for the isospin $I$ state at an energy of the rest mass of $\ko $. $\omega $ is a measure of deviation from the $\De I=1/2$ rule, and may be inferred, e.g., from 
%
%
\begin{eqnarray}
r&\equiv &\frac{\ga _S(\pi ^+\pi ^-)-2\ga _S(\pi ^0\pi ^0)}{\ga _S(\pi ^+\pi ^-)+\ga _S(\pi ^0\pi ^0)} \no \\ 
&=& \frac{4\re (\omega ')-2|\omega '|^2}{1+2|\omega '|^2} ~.
\end{eqnarray}
Here and in the following, $\ga _{S,L}(n)$ denotes the partial width for $K_{S,L}$ to decay into the final state $|n\ket $.

\subsection{$3\pi$ and $\pi^+\pi^-\ga$ modes }
\label{sec:3.2}
The experimentally measured quantities are 
%
%
\begin{subequations}
\begin{equation}
\eta _{+-0}= \frac{\bra \pi ^+\pi ^-\pi ^0,\mbox{outgoing}|\Hw |K_S\ket }{\bra \pi ^+\pi ^-\pi ^0,\mbox{outgoing}|\Hw |K_L\ket }~, 
\end{equation}
\begin{equation}
\eta _{000}= \frac{\bra \pi ^0\pi ^0\pi ^0,\mbox{outgoing}|\Hw |K_S\ket }{\bra \pi ^0\pi ^0\pi ^0,\mbox{outgoing}|\Hw |K_L\ket }~,
\end{equation}
\end{subequations}
%
%
\begin{equation}
\eta _{+-\ga }= \frac{\bra \pi ^+\pi ^-\ga ,\mbox{outgoing}|\Hw |K_L\ket }{\bra \pi ^+\pi ^-\ga ,\mbox{outgoing}|\Hw |K_S\ket }~.
\end{equation}
We shall treat the $3\pi $ $(\pi ^+\pi ^-\ga )$ states as purely $CP$-odd ($CP$-even). 
\subsection{Semi-leptonic modes}
\label{sec:3.3}
The final states of particular interest are $|\ell^+\ket \equiv |\pi^-\ell^+\nu_{\ell}\ket $ and $|\ell^-\ket \equiv |\pi^+\ell^- \nub_{\ell}\ket$, where $\ell=e$ or $\mu$, and the well measured time-independent asymmetry parameter related to $CP$ violation is
%
%
\begin{equation}
d^{\ell }_L = \frac{\ga _L(\pi ^-\ell ^+\nu _{\ell })-\ga _L(\pi ^+\ell ^-\nub _{\ell })}{\ga _L(\pi ^-\ell ^+\nu _{\ell })+\ga _L(\pi ^+\ell ^-\nub _{\ell })}~.
\end{equation}
CPLEAR Collaboration [14-16] have furthermore defined and measured two kinds of time-dependent experimental asymmetry parameters $A_T^{exp}(t)$ and $A_{\de}^{exp}(t)$ which are related to
%
%
\begin{subequations}
\begin{equation}
d^{\ell}_1(t)=\frac{|\bra \ell^+|\Hw|\kob(t)\ket|^2-|\bra \ell^-|\Hw|\ko(t)\ket|^2}{|\bra \ell^+|\Hw|\kob(t)\ket|^2+|\bra \ell^-|\Hw|\ko(t)\ket|^2}~,
\end{equation}
\begin{equation}
d^{\ell}_2(t)=\frac{|\bra \ell^-|\Hw|\kob(t)\ket|^2-|\bra \ell^+|\Hw|\ko(t)\ket|^2}{|\bra \ell^-|\Hw|\kob(t)\ket|^2+|\bra \ell^+|\Hw|\ko(t)\ket|^2}~.
\end{equation}
\end{subequations}
%
%
\section{Parametrization and conditions imposed by $CP$, $T$ and $CPT$ symmetries}
We shall parametrize the ratios of the mixing parameters, $q_S/p_S$ and $q_L/p_L$, as
%
%
\begin{equation}
\left. 
	\begin{array}{c}
	\frac{\dps{q_S}}{\dps{p_S}}=e^{i\al _K}\frac{\dps{1-\e _S}}{\dps{1+\e _S}}~, \\ \\
	\frac{\dps{q_L}}{\dps{p_L}}=e^{i\al _K}\frac{\dps{1-\e _L}}{\dps{1+\e _L}}~,
	\end{array}
\right. 
\end{equation}
and $\e _{S,L}$ further as 
%
%
\begin{equation}
\e _{S,L}=\e \pm \de ~.
\end{equation}
With the aid of Eqs.(2.1) and (2.8), one sees that $CP$, $T$ and $CPT$ symmetries impose such coditions on $\e$ and $\de$ as
%
%
\begin{subequations}
\begin{eqnarray}
CP  &\to& \e = 0~, \qquad \de = 0~; \\
T   &\to& \e = 0~; \\
CPT &\to& \de = 0~.
\end{eqnarray} 
\end{subequations}
Since $CP$ violation is known to be very tiny (see below), one may treat $\e$ and $\de$ as small parameters. From Eqs.(2.7), (2.8) and (2.9), one then derive [4]
%
%
\begin{subequations}
\begin{eqnarray}
\De m &\simeq& 2\re(M_{12}e^{i\alpha_K})~, \\
\De \ga &\simeq& 2\re(\Gamma_{12}e^{i\alpha_K})~,
\end{eqnarray}
\end{subequations}
%
%
\begin{subequations}
\begin{eqnarray}
\e &\simeq& (\La_{12}e^{i\alpha_K}-\La_{21}e^{-i\alpha_K})/2\De \la~, \\
\de &\simeq& (\La_{11}-\La_{22})/2\De \la~,
\end{eqnarray}
\end{subequations}
from which it follows that [4,5]
%
%
{
\setcounter{enumi}{\value{equation}}
\addtocounter{enumi}{1}
\setcounter{equation}{0}
\renewcommand{\theequation}{\thesection.\theenumi\alph{equation}}
\begin{equation}
\e _{\| }\equiv \re [\e \exp (-i\phi _{SW})]\simeq \frac{-2\im (M_{12}e^{i\al _K})}{\sqrt{(\ga _S-\ga _L)^2+4(\De m)^2}}~,
\end{equation}
\begin{equation}
\e _{\perp }\equiv \im [\e \exp (-i\phi _{SW})]\simeq \frac{\im (\Gamma _{12}e^{i\al _K})}{\sqrt{(\ga _S-\ga _L)^2+4(\De m)^2}}~,
\end{equation}
\setcounter{equation}{\value{enumi}}}%
%
%
{
\setcounter{enumi}{\value{equation}}
\addtocounter{enumi}{1}
\setcounter{equation}{0}
\renewcommand{\theequation}{\thesection.\theenumi\alph{equation}}
\begin{equation}
\de _{\|}\equiv \re [\de \exp (-i\phi _{SW})]\simeq \frac{\Gamma _{11}-\Gamma _{22}}{2\sqrt{(\ga _S-\ga _L)^2+4(\De m)^2}}~,
\end{equation}
\begin{equation}
\de _{\perp }\equiv \im [\de \exp (-i\phi _{SW})]\simeq \frac{M_{11}-M_{22}}{\sqrt{(\ga _S-\ga _L)^2+4(\De m)^2}}~,
\end{equation}
\setcounter{equation}{\value{enumi}}}%
where 
%
%
\begin{subequations}
\begin{eqnarray}
\De m \equiv m_S-m_L~,~~  \De \ga &\equiv& \ga_S-\ga_L~, ~~ \De \la \equiv \la_S-\la_L~, \\ 
\phi _{SW} &\equiv& \tan ^{-1}\left( \frac{-2\De m}{\De \ga} \right)~. 
\end{eqnarray}
\end{subequations}
$\phi _{SW}$ is often called the superweak phase. 

Paying particular attention to the $2\pi$ and semi-leptonic decay modes, we shall  parametrize amplitudes for $|\ko\ket$ and $|\kob\ket$ to decay into $|(2\pi)_I\ket$,
%
%
\begin{subequations}
\begin{eqnarray}
	A_I &\equiv& \bra(2\pi)_I|\Hw|\ko\ket ~, \\
	\bar{A}_I &\equiv& \bra(2\pi)_I|\Hw|\kob\ket ~,
\end{eqnarray}
\end{subequations}
as
%
%
\begin{subequations}
\begin{eqnarray}
	A_I &=& F_I(1 + \e_I)e^{i(\phi_I + \theta_I + \al_K/2)}~, \\
	\overline{A}_I &=& F_I(1 - \e_I)e^{i(\phi_I - \theta_I - \al_K/2)}~,
\end{eqnarray}
\end{subequations}
and amplitudes for $|\ko\ket$ and $|\kob\ket$ to decay into $|\ell^+\ket$ or $|\ell^-
\ket$, 
%
%
\begin{subequations}
\begin{eqnarray}
	A_{\ell +} &\equiv& \bra\ell^+|\Hw|\ko\ket ~, \\ 
	\overline{A}_{\ell -} &\equiv& \bra\ell^-|\Hw|\kob\ket ~, \\ 
	\overline{A}_{\ell +} &\equiv& \bra\ell^+|\Hw|\kob\ket ~, \\
	A_{\ell -} &\equiv& \bra\ell^-|\Hw|\ko\ket ~,
\end{eqnarray}
\end{subequations}
as
%
%
\begin{subequations}
\begin{eqnarray}
	A_{\ell +} &=& F_{\ell}(1 + \e_{\ell})e^{i(\phi_{\ell} + \theta_{\ell} + \al_K/2)}~, \\ 
	\overline{A}_{\ell -} &=& F_{\ell}(1 - \e_{\ell})e^{i(\phi_{\ell} - \theta_{\ell} - \al_K/2)}~, \\ 
	\overline{A}_{\ell +} &=& x_{\ell +}F_{\ell}(1 + \e_{\ell})e^{i(\phi_{\ell} + \theta_{\ell} - \al_K/2)}~, \\
	A_{\ell -} &=& x_{\ell -}^*F_{\ell}(1 - \e_{\ell})e^{i(\phi_{\ell} - \theta_{\ell} + \al_K/2)}~.
\end{eqnarray}
\end{subequations}
Here, $F_I$ and $F_{\ell}$ are real and positive, $\e_I$, $\phi_I$, $\theta_I$, $\e_{\ell}$, $\phi_{\ell}$ and $\theta_{\ell}$ are real, while $x_{\ell+}$ and $x_{\ell-}$ are 
complex in general.  $x_{\ell+}$ and $x_{\ell-}$, which measure violation of the $\De S=\De Q$ rule, will further be parametrized as
%
%
\begin{equation}
x_{\ell+} = x_{\ell}^{(+)} + x_{\ell}^{(-)}~, \qquad x_{\ell-} = x_{\ell}^{(+)} - x_{\ell}^{(-)}~.
\end{equation}
Note that we have defined our amplitude parameters $F_I$, $\e_I$, $\phi_I$, $\theta_I$, $F_{\ell}$, $\e_{\ell}$, $\phi_{\ell}$, $\theta_{\ell}$, $x_{\ell}^{(+)}$ and $x_{\ell}^{(-)}$, and our mixing parameters $\e$ and $\de$ as well, in such a way that they are all invariant with respect to rephasing of $|\ko\ket$ and $|\kob\ket$,
%
%
\begin{equation}
|\ko \ket \to |\ko \ket '=|\ko \ket e^{-i\xi _K}~, ~~ |\kob \ket \to |\kob \ket '=|\kob \ket e^{i\xi _K}~,
\end{equation}
in spite that $\al_K$ itself is not invariant with respect to this rephasing [4,21,22].\footnote{In contrast, $\beta_K$ is invariant with respect to the rephasing (4.14).}

As will be shown explicitly in Appendix, one may convince himself that phase ambiguities associated with $|(2\pi )_I\ket$, $|\ell^+\ket$ and $|\ell^-\ket$ allow one, without loss of generality, to take
%
%
\begin{equation}
\phi_I = 0~, \qquad \phi_{\ell} = 0~, \qquad \theta_{\ell} = 0~,
\end{equation}
and that $CP$, $T$ and $CPT$ symmetries impose such conditions as
%
%
\begin{subequations}
\begin{eqnarray}
CP &\to& \e_I = 0~, \qquad \theta_I = 0~, \qquad \e_{\ell} = 0~, \no \\
   &   & \im(x_{\ell}^{(+)}) = 0~, \qquad \re(x_{\ell}^{(-)}) = 0~; \\ 
T  &\to& \theta_I = 0~, \no \\
   &   & \im(x_{\ell}^{(+)}) = 0~, \qquad \im(x_{\ell}^{(-)}) = 0~; \\ 
CPT&\to& \e_I = 0~, \qquad \e_{\ell} = 0~, \no \\
   &   & \re(x_{\ell}^{(-)}) = 0~, \qquad \im(x_{\ell}^{(-)}) = 0~. 
\end{eqnarray}
\end{subequations}
%
%
\section{Formulas relevant for numerical analyses }
We shall adopt a phase convention which gives Eq.(4.15). Observed or expected smallness of violation of $CP$, $T$ and $CPT$ symmetries and of the $\De I = 1/2$ and $\De Q = \De S$ rules allows us to treat all our parameters, $\e$, $\de$, $\e_I$, $\theta_I$, $\e_{\ell}$, $x_{\ell}^{(+)}$, $x_{\ell}^{(-)}$ as well as $\omega'$ as small, and, from Eqs.(3.2), (3.3), (3.4a,b), (3.6) and (3.9), one finds, to the leading order in these small parameters,
%
%
\begin{equation}
\omega \simeq F_2/F_0~,
\end{equation}
%
%
\begin{equation}
\eta_I \simeq \e -\de + \e_I + i\theta_I~,
\end{equation}
%
%
\begin{subequations}
\begin{eqnarray}
\eta_{+-} &\simeq& \eta_0 + \e'~, \\
\eta_{00} &\simeq& \eta_0 - 2\e'~,
\end{eqnarray}
\end{subequations}
%
%
\begin{equation}
r \simeq 4\re(\omega')~,
\end{equation}
%
%
\begin{equation}
d_L^{\ell} \simeq 2(\re(\e-\de)+\e_{\ell}-\re(x_{\ell}^{(-)}))~,
\end{equation}
where
%
%
\begin{equation}
\e' \equiv (\eta_2-\eta_0)\omega'~.
\end{equation}
The time-dependent asymmetry parameters $d_1^{\ell}(t)$ and $d_2^{\ell}(t)$, defined by Eqs.(3.10a,b), behave as
%
%
\begin{subequations}
\begin{eqnarray}
d_1^{\ell}(t \gg \tau_S) \simeq 4\re(\e)+2(\e_{\ell}-\re(x_{\ell}^{(-)}))~, \\
d_2^{\ell}(t \gg \tau_S) \simeq 4\re(\de)-2(\e_{\ell}-\re(x_{\ell}^{(-)}))~,
\end{eqnarray}
\end{subequations}
while the experimental time-dependent asymmetry parameters $A_T^{exp}(t)$ and $A_{\de}^{exp}(t)$ defined and measured by CPLEAR Collaboration [14-16,20] behave as
%
%
\begin{subequations}
\begin{eqnarray}
A_T^{exp}(t \gg \tau_S) &\simeq& 4(\re(\e)+\e_{\ell}-\re(x_{\ell}^{(-)}))~, \\
A_{\de}^{exp}(t \gg \tau_S) &\simeq& 8\re(\de)~.
\end{eqnarray}
\end{subequations}
%
%
%
%

From Eqs.(5.3a,b), it follows that
%
%
\begin{equation}
\eta_0 \simeq (2/3)\eta_{+-}(1+(1/2)|\eta_{00}/\eta_{+-}|e^{i\De\phi})~,
\end{equation}
and, treating $|\e'/\eta_0|$ as a small quantity, which is justifiable empirically (see below), one further obtains
%
%
\begin{equation}
\eta _{00}/\eta _{+-} \simeq 1 - 3\e'/\eta_0~,
\end{equation}
or
%
%
\begin{subequations}
\begin{eqnarray}
\re(\e'/\eta_0) &\simeq& (1/6)(1 - |\eta_{00}/\eta_{+-}|^2)~, \\
\im(\e'/\eta_0) &\simeq& -(1/3)\De\phi~,
\end{eqnarray}
\end{subequations}
where
%
%
\begin{equation}
\De\phi \equiv \phi_{00} - \phi_{+-}~.
\end{equation}
On the other hand, from Eqs.(3.5), (5.1), (5.2) and (5.6), $\e'/\eta_0$ may be related to $\e_2-\e_0$ and $\theta_2-\theta_0$:
%
%
\begin{equation}
\e'/\eta_0 = -i\re(\omega')((\e_2-\e_0)+i(\theta_2-\theta_0))e^{-i\De\phi'}/[|\eta_0|\cos(\de_2-\de_0)]~, 
\end{equation}
where
%
%
\begin{equation}
\De\phi' \equiv \phi_0 - \de_2 + \de_0 - \pi/2~.
\end{equation}
Also, from Eqs.(5.2) and (5.5), one may derive
%
%
\begin{equation}
\re(\eta_0)-d_L^{\ell}/2 = \e_0-\e_{\ell}+\re(x_{\ell}^{(-)})~.
\end{equation}

Furthermore, noting that 
\[
\bra K_S|K_L\ket \simeq 2[\re(\e) - i\im(\de)]~,
\]
one may use the Bell-Steinberger relation, Eq.(2.10), to express $\re(\e)$ and $\im(\de)$ in terms of measured quantities. By taking $2\pi $, $3\pi $, $\pi ^+\pi ^-\ga $ and $\pi \ell \nu_{\ell }$ intermediate states into account in Eq.(2.11) and making use of the fact $\ga _S \gg \ga _L$, we derive
%
%
\begin{eqnarray}
\re (\e ) & \simeq & \frac{1}{\sqrt{\ga _S^2+4(\De m)^2}}\times \no \\ 
 &&  \Big{[} ~\ga _S(\pi ^+\pi ^-)|\eta _{+-}|\cos (\phi _{+-}-\phi _{SW}) \no \\ 
 && ~~+\ga _S(\pi ^0\pi ^0)|\eta _{00}|\cos (\phi _{00}-\phi _{SW}) \no \\ 
 && ~~+\ga _S(\pi ^+\pi ^-\ga )|\eta _{+-\ga }|\cos (\phi _{+-\ga }-\phi _{SW}) \no \\ 
 && ~~+\ga _L(\pi ^+\pi ^-\pi ^0)\{ \re (\eta _{+-0})\cos \phi _{SW} -\im (\eta _{+-0})\sin \phi _{SW}\} \no \\ 
 && ~~+\ga _L(\pi ^0\pi ^0\pi ^0)\{ \re (\eta _{000})\cos \phi _{SW} -\im (\eta _{000})\sin \phi _{SW}\} \no \\ 
 && ~~+2\sum _{\ell }\ga _L(\pi \ell \nu _{\ell })\{\e_{\ell}\cos \phi _{SW} -\im(x_{\ell}^{(+)})\sin \phi _{SW}\} \Big{]}, 
\end{eqnarray}
%
%
\begin{eqnarray}
\im (\de ) & \simeq & \frac{1}{\sqrt{\ga _S^2+4(\De m)^2}}\times \no \\ 
 &&  \Big{[} ~-\ga _S(\pi ^+\pi ^-)|\eta _{+-}|\sin (\phi _{+-}-\phi _{SW}) \no \\ 
 && ~~-\ga _S(\pi ^0\pi ^0)|\eta _{00}|\sin (\phi _{00}-\phi _{SW}) \no \\ 
 && ~~-\ga _S(\pi ^+\pi ^-\ga )|\eta _{+-\ga }|\sin (\phi _{+-\ga }-\phi _{SW}) \no \\ 
 && ~~+\ga _L(\pi ^+\pi ^-\pi ^0)\{ \re (\eta _{+-0})\sin \phi _{SW} +\im (\eta _{+-0})\cos \phi _{SW}\} \no \\ 
 && ~~+\ga _L(\pi ^0\pi ^0\pi ^0)\{ \re (\eta _{000})\sin \phi _{SW} +\im (\eta _{000})\cos \phi _{SW}\} \no \\ 
 && ~~+2\sum _{\ell }\ga _L(\pi \ell \nu _{\ell })\{\e_{\ell}\sin \phi _{SW} +\im(x_{\ell}^{(+)})\cos \phi _{SW}\} \Big{]}.
\end{eqnarray}
If, however, one retains the contribution of the $2\pi$ intermediate states alone, the 
Bell-Steinberger relation gives 
simply
%
%
\begin{equation}
\re(\e) - i\im(\de) \simeq |\eta_0|e^{i\De\phi"}\cos\phi_{SW}~,
\end{equation}
where
%
%
\begin{equation}
\De\phi" \equiv \phi_0 - \phi_{SW}~.
\end{equation}
Note that exactly the same equation as Eq.(5.18) can be derived from Eqs.(4.6b) and (4.7a).
%
%
\section{Evaluation of the symmetry-violating parameters}
%
%
%
\begin{table}
  \caption{Inputs}
  \label{tab:1}
 \begin{center}
  \begin{tabular}{c|c|c|c|c}
   \hline \hline 
& Quantity  & Value & Unit & Ref.  \\ \hline 
       & $\tau_S$ & 0.8959 $\pm$ 0.0004 & $10^{-10}s$ & [17, 18] \\ 
       & $\tau_L$ & 5.17 $\pm$ 0.04 & $10^{-8}s$ & [19] \\ 
       & $-\Delta m$ & 0.5284$\pm$0.0011 & $10^{10}s^{-1}$ & [17, 18]  \\ \hline 
$2\pi$ & $\gamma_S(\pi^{+} \pi^{-})/\gamma_S$ & 68.60$\pm$0.27 & $\%$ & [19]
   \\
       & $\gamma_S(\pi^{0} \pi^{0})/\gamma_S$ & 31.40$\pm$0.27 & $\%$ & [19] 
   \\ 
       & $\delta_2-\delta_0 $ & -42 $\pm$ 20 & $^\circ$  &  \\
       & $|\eta _{+-}|$ & $2.286 \pm 0.017$ & $10^{-3}$ & [19] \\
       & $\phi_{+-}$ & 43.4$\pm$0.7 & $^\circ$ & [19] \\
       & $|\eta_{00}/\eta_{+-}|$ & $0.99484\pm 0.00054$ & & [17, 18] \\
       & $\Delta \phi $ & 0.22$\pm$0.45 & $ ^\circ$ & [17, 18] \\ \hline 

$\pi^{+} \pi^{-} \gamma$ & $\gamma_S(\pi^{+} \pi^{-} \gamma)/\gamma_S$
   & $0.178 \pm 0.005$ & $\%$ & [19] \\ 
          & $|\eta_{+-\gamma}|$ & $2.35 \pm 0.07$ &
   $10^{-3}$ & [19]  \\ 
          & $\phi_{+-\gamma}$ & $44 \pm 4$ & $^\circ$ & [19] \\ \hline 
   $3\pi$ & $\gamma_L(\pi^{+} \pi^{-} \pi^{0})/\gamma_L$ & $12.58 \pm 0.19$ & $\%$ & [19] \\
          & $\gamma_L(\pi^{0} \pi^{0} \pi^{0})/\gamma_L$ & $21.08 \pm 0.27$ & $\%$ & [19] \\ 
          & $\re(\eta_{+-0})$ & $-0.002 \pm 0.008$ & & [13, 16] \\
          & $\im(\eta_{+-0})$ & $-0.002 \pm 0.009$ & & [13, 16] \\ 
          & $\re(\eta_{000})$ & $0.08 \pm 0.11$ & & [13, 16] \\ 
          & $\im(\eta_{000})$ & $0.07 \pm 0.16$ & & [13, 16] \\ \hline 
          
$\pi \ell \nu$ & $d^{\ell}_{L}$ & 3.31$\pm$0.06 & $10^{-3}$ & [17, 18]  \\ 
               & $\sum_{\ell}\gamma_L(\pi\ell\nu)/\gamma_L$ & $65.97 \pm 0.30$ & $\%$& [19] \\ 
               & $\re(\de)$ & $3.0\pm 3.4$ & $10^{-4}$ & [20] \\
               & $\re(x_{\ell}^{(-)})$ & 0.2$\pm $1.3 & $10^{-2}$ & [20] \\
               & $\im(x^{(+)}_{\ell})$ & $1.2 \pm 2.2$ & $10^{-2}$ & [20] \\ \hline 
  \end{tabular}
 \end{center}
\end{table}
The data used as inputs in the numerical analyses given below are tabulated in Table 1. 
Many of them are from Particle Data Group [19] and some are the new world averages 
reported by KTeV Collaboration [17] and NA48 Collaboration [18]. As for the 
values of $\eta_{+-0}$ and $\eta_{000}$, we use those obtained without recourse to $CPT$ 
symmetry by CPLEAR Collaboration [13,16]. This Collaboration further succeeded, from 
an unconstrained fit to the experimental time-dependent asymmetry $A_{\de}^{exp}(t)$ they defined and measured, in determining $\im(\de)$, $\re(\de)$, $\im(x_{\ell}^{(+)})$ and 
$\re(x_{\ell}^{(-)})$ simultaneously [15,20]:
%
%
\begin{subequations}
\begin{eqnarray}
\im(\de) &=& (-1.5 \pm 2.3) \times 10^{-2}~, \\
\re(\de) &=& (~3.0 \pm 3.4) \times 10^{-4}~, \\
\im(x_{\ell}^{(+)}) &=& (1.2 \pm 2.2) \times 10^{-2}~, \\
\re(x_{\ell}^{(-)}) &=& (0.2 \pm 1.3) \times 10^{-2}~.
\end{eqnarray}
\end{subequations}
We shall include Eqs.(6.1b,c,d) in our list of the input data. As for the value of 
$\de_2-\de_0$, we shall use the value obtained by Chell and Olsson [26], with the error 
extended arbitrarily by a factor of five to take account of its possible uncertainty [27].

Our analysis consists of three steps:

{\it The first step.}~~We use Eq.(4.8b) to find $\phi_{SW}$ from $\De m$ 
and $\ga_S$, use Eqs.(3.6) and (5.4) to find $\re(\omega')$ from 
$\ga_S(\pi^+\pi^-)/\ga_S$ and $\ga_S(\pi^0\pi^0)/\ga_S$, and 
further use Eqs.(5.11a,b) and (5.9) to find 
$\re(\e'/\eta_0)$, $\im(\e'/\eta_0)$, $|\eta_0|$ and $\phi_0$ from 
$|\eta_{00}/\eta_{+-}|$, $\De \phi$, $|\eta_{+-}|$ and $\phi_{+-}$. These 
results are shown as the intermediate outputs in Table 2.

{\it The second step.}~~The values of $\eta_0$, $\e'/\eta_0$, $\phi_{SW}$ and 
$\re(\omega')$ obtained, supplemented with the value of $\de_2-\de_0$, are used as inputs to find $\theta_2-\theta_0$ and $\e_2-\e_0$ with the help of Eqs.(5.13) and (5.14). 
The value of $\re(\eta_0)$ is combined with that of $d_L^{\ell}$ to find $\e_0-\e_{\ell}+\re(x_{\ell}^{(-)})$ with the aid of Eq.(5.15). The value of $\re(\de)$ is combined with 
the value of $\re(\eta_0)$ to find $\re(\e)+\e_0$ with the aid of Eq.(5.2) and combined 
with the value of $d_L^{\ell}$ to find $\re(\e)+\e_{\ell}-\re(x_{\ell}^{(-)})$ with the 
aid of Eq.(5.5). These results are shown in Table 3.

{\it The third step}~~Eqs. (5.5) and (5.16) are solved, with $d_L^{\ell}$, $\re(\de)$, $\re(x_{\ell}^{(-)})$ and $\im(x_{\ell}^{(+)})$ regarded as known, to find $\re(\e)$ and $\e_{\ell}$, and Eq.(5.17) is then used to find $\im(\de)$. Finally, the values of 
$\re(\e)$, $\re(\de)$ and $\im(\de)$ are combined with the value of $\eta_0$ to determine or constrain $\e_0$ and $\im(\e)+\theta_0$ with the help of Eq.(5.2). These results are 
also shown in Table 3.

In relation to the third step of our analysis, which relies on the use of the 
Bell-Steinberger relation, Eqs.(5.16) and (5.17), a couple of remarks are in order:

(1) The large uncertainty associated with $\e_{\ell}$ comes primarily from that of $\re(x_{\ell}^{(-)})$, while, thanks to the fact that $2\pi$ decay modes dominate over all the 
other decay modes, the large uncertainty associated with $\re(x_{\ell}^{(-)})$ as well as $\im(x_{\ell}^{(+)})$ has little influence on determination of $\re(\e)$ and $\im(\de)$.

(2) To see how much the decay modes other than $2\pi$ modes contribute, we perform a 
similar analysis with the aid of the simplified version (i.e., $2\pi$ dominance version) 
of the Bell-Steinberger relation, Eq.(5.18), to obtain the result shown in Table 4.

(3) If one is not willing to rely on the Bell-Steinberger relation at all, one will not be able to constrain $\re(\e)$, $\e_0$ and $\e_{\ell}$ separately and hence will not be able to establish $\re(\e) \neq 0$.\footnote{From measurement of the other experimental 
time-dependent asymmetry parameter $A_T^{exp}(t)$, CPLEAR Collaboration found [14,20]
%
%
\begin{subequations}
\begin{eqnarray}
\re(\e) &=& (1.55 \pm 0.43) \times 10^{-3}~, \\
\im(x_{\ell}^{(+)}) &=& (1.2~ \pm 2.1~) \times 10^{-3}~.
\end{eqnarray}
\end{subequations}
However, since the value of $\re(\e)$ shown here is determined predominantly by the 
asymptotic value of $A_T^{exp}(t)$ under the assumption of $CPT$ symmetry (spesifically, 
$\e_{\ell} = 0$ and $\re(x_{\ell}^{(-)}) = 0$), it seems that the result (6.2a) is better to be interpreted as giving (see Eq.(5.8a)) 
%
%
\begin{equation}
\re(\e)+\e_{\ell}-\re(x_{\ell}^{(-)}) = (1.55 \pm 0.43) \times 10^{-3}~,
\end{equation}
and therefore that $\re(\e)$ remains unconstrained. In our previous analyses [9], we have accepted the assumption $\re(x_{\ell}^{(-)})=0$ and used Eqs.(6.2b) and (6.3) as a part of our inputs.} One may, instead, use the CPLEAR result on $\im(\de)$, Eq.(6.1a), as one of the inputs to derive
%
%
\begin{equation}
\im(\e)+\theta_0 = (-1.34 \pm 2.30) \times 10^{-2}~.
\end{equation}
%
%
%
%
\begin{table}
 \caption{Intermediate outputs}
\label{tab:2}
\begin{center}
 \begin{tabular}{c|c|c}
  \hline \hline
  Quantity & Result & Unit  \\ \hline 
  $\phi_{SW}$ & $43.48 \pm 0.06$ & $^\circ$ \\ 
  $\re(\omega ')$ & $1.450 \pm 0.151$ & $10^{-2}$ \\ 
  $\re(\e'/\eta_0)$ & $1.72 \pm 0.18$ & $10^{-3}$ \\ 
  $\im(\e'/\eta_0)$ & $-1.28 \pm 2.62$ & $10^{-3}$ \\ 
  $|\eta_0|$ & $2.282 \pm 0.017$ & $10^{-3}$ \\  
  $\phi_0$ & $43.47 \pm 0.96$ & $^\circ$ \\ \hline 
 \end{tabular}
\end{center}
\end{table}
%
%
\begin{table}
\caption{Constraints (in unit of $10^{-3}$) to $CP$, $T$ and/or $CPT$-violating parameters .}
\label{tab:3}
\begin{center}
		\begin{tabular}{c|c}
		\hline \hline 
		 Quantity & Result  \\ \hline 

         $\theta_2-\theta_0 $  & $0.189\pm 0.091$ \\
         $\e_2-\e_0 $  & $0.165\pm 0.317$ \\
         $\e_0-\e_{\ell}+\re(x_{\ell}^{(-)}) $  & $0.001\pm 0.042$ \\ 
         $\re(\e)+\e_0$  & $1.956\pm 0.344$ \\ 
         $\re(\e)+\e_{\ell}-\re(x_{\ell}^{(-)}) $  & $1.955\pm 0.341$ \\ \hline 
                  
		 $\re(\e) $  & $1.652\pm 0.048$    \\ 
		 $\im(\de) $  & $0.045\pm 0.050$    \\ 
		 $\e_{\ell}-\re(x_{\ell}^{(-)}) $ & $0.310\pm 0.344$   \\
		 $\e_{\ell} $  & $2.303\pm 13.005$   \\ 
		 
		 $\e_0 $  & $0.304\pm 0.345$  \\ 
		 $\im(\e) + \theta_0 $  & $1.615\pm 0.059$  \\ \hline
		\end{tabular}
\end{center}
\end{table}
%
%
\begin{table}
\caption{Constraints (in unit of $10^{-3}$) to $CP$, $T$ and/or $CPT$-violating parameters obtained with the aid of the simplified version of the Bell-Steinberger relation (5.18).}
\label{tab:4}
\begin{center}
		\begin{tabular}{c|c}
		\hline \hline 
		 Quantity & Result  \\ \hline
         
		 $\re(\e) $  & $1.656\pm 0.012 $ \\ 
		 $\im(\de) $  & $0.0003\pm 0.028 $ \\ 
         $\e_{\ell}-\re(x_{\ell}^{(-)}) $  & $0.299\pm 0.342 $ \\ 
		 $\e_{\ell} $  & $2.300\pm 13.004 $ \\ 

		 $\e_0 $  & $0.300\pm 0.341 $ \\ 
		 $\im(\e) + \theta_0 $  & $1.570\pm 0.012 $ \\ \hline
		\end{tabular}
\end{center}
\end{table}
%
%
%
\begin{table}
\caption{Comparison of our results with the CPLEAR result obtained with the Bell-Steinberger relation taken as a constraint (The values underlined are inputs; all in unit of $10^{-3}$).}
\label{tab:5}
\begin{center}
		\begin{tabular}{l|cc|c}
		\hline \hline 
 & Our result & Our result & CPLEAR  \\ 
 & & $2\pi$ dominance & Ref. [16]  \\ \hline 
$\re(\de)$ & \underline{$0.30\pm 0.34$} & \underline{$0.30\pm 0.34$} & $0.24\pm 0.28$ \\ 
$\im(x_{\ell}^{(+)})$ & \underline{$12\pm 22$} & --- & $-2.0\pm 2.7$ \\ 
$\re(x_{\ell}^{(-)})$ & \underline{$2\pm 13$} & \underline{$2\pm 13$} & $-0.5\pm 3.0$ \\ 
$\re(\e)$ & $1.652\pm 0.048$ & $1.656\pm 0.012$ & $1.649\pm 0.025$ \\ 
$\im(\delta)$ & $0.045\pm 0.050$ & $0.0003\pm 0.028$ & $0.024\pm 0.050$ \\ 
$\e_{\ell}-\re(x_{\ell}^{(-)})$ & $0.310\pm 0.344$ & $0.299\pm 0.342$ & $0.2\pm 0.3$ \\
$\e_{\ell}$ & $2.303\pm 13.005$ & $2.300\pm 13.004$ & $-0.3\pm 3.1$ \\ \hline
   		\end{tabular}
\end{center}
\end{table}
\vspace{-0.8cm} 
%
%
\section{Summary and concluding remarks}
In order to identify or search for possible violation of $CP$,~$T$ and $CPT$ symmetries in the $\ko-\kob$ system, parametrizing the mixing parameters and the relevant decay 
amplitudes in a convenient and well-defined way, we have, with the relevant experimental 
data used as inputs and partly with the aid of the Bell-Steinberger relation, performed 
numerical analyses to determine or constrain the symmetry-violating parameters separately as far as possible.

The numerical outputs of our analyses are shown in Table 2, Table 3 and Table 4 (see also Table 5), and the main results may be summarized as follows:

(1) The $2\pi$ data directly give $\theta_2-\theta_0 = (1.89 \pm 0.91) \times 
10^{-4}$ and $\e_2-\e_0 = (1.65 \pm 3.17) \times 10^{-4}$ , where possible large 
uncertainty associated with $\de_2-\de_0$ has been fully taken into account. These results indicate that $CP$ and $T$ symmetries are definitively violated at least at the level of $10^{-4}$, while $CPT$ symmetry holds presumably down to the same level, in decay of 
$\ko$ and $\kob$ into $2\pi$ states.

(2) The well-measured leptonic asymmetry $d_L^{\ell}$, combined with the $2\pi$ data, 
gives $\e_0-\e_{\ell}+\re(x_{\ell}^{(-)}) = (0.1 \pm 4.2) \times 10^{-5}$, which is a 
tight constraint to a particular combination of the direct $CP$ and $CPT$ violating 
parameters.

(3) The data on $2\pi$ modes and on $d_L^{\ell}$, combined with the CPLEAR value of $\re(\de)$, give $\re(\e)+\e_0 = (1.956 \pm 0.344) \times 10^{-3}$ and $\re(\e)+\e_{\ell}-\re(x_{\ell}^{(-)}) = (1.955 \pm 0.341) \times 10^{-3}$ respectively, which implies 
that $CP$ and $T$ symmetries are violated at the level of $10^{-3}$ in the $\ko-\kob$ 
mixing and/or that $CP$ and $CPT$ symmetries are violated at the same level in $2\pi$ and $\pi\ell\nu_{\ell}$ decays.

(4) The use of the CPLEAR results on $\re(\de)$, $\re(x_{\ell}^{(-)})$ and $\im(x_{\ell}^{(+)})$ as a part of the inputs, aided by the Bell-Steinberger relation, 
enables one to determine or constrain many of the remaining symmetry-violating 
parameters separately. Remarkably, $\re(\e)$ and $\im(\de)$ turn out to be rather well 
determined or constrained: $\re(\e) = (1.652 \pm 0.048) \times 10^{-3}$, $\im(\de) = (4.5 \pm 5.0) \times 10^{-5}$. The former (latter) indicates that $CP$ and $T$ symmetries are violated at the level of $10^{-3}$ ($CPT$ symmetry holds nearly down to the level of $10^{-5}$) in mixing of $\ko$ and $\kob$.

(5) The results mentioned in (4) above in turn give $\e_0 = (3.04 \pm 3.45) \times 10^{-4}$ and $\im(\e)+\theta_0 = (1.615 \pm 0.059)\times 10^{-3}$. The latter implies that $CP$ and $T$ symmetries are violated definitively at least at the level of $10^{-3}$ in the $\ko-\kob$ mixing and/or in $2\pi$ decays.\footnote{Without the aid of Bell-Steinberger relation, $\im(\e)+\theta_0$ remains ill-constrained (see Eq.(6.4)). Note also that we are not able to separate $\theta_0$ from $\im(\e)$ (see Remarks (4) and (5) in Appendix for related discussion).}

In the present work, in order to be as phenomenological and comprehensive as possible, we have chosen Eqs.(6.1b,c,d), those CPLEAR results which were obtained without any 
constraint, as a part of our inputs.\footnote{In addition to the analyses leading to the 
results shown in Eqs.(6.1a,b,c,d) and in Eqs.(6.2a,b), CPLEAR Collaboration [16] 
further carried out a simultaneous fit to both $A_T^{exp}(t)$ and $A_{\de}^{exp}(t)$, with the Bell-Steinberger relation taken as a constraint, and succeeded in determining 
$\re(\e)$, $\im(\de)$, $\re(\de)$, $\im(x_{\ell}^{(+)})$, $\re(x_{\ell}^{(-)})$ and 
$\re(\e_{\ell})$ simultaneously. These results, also shown in Table 5 for comparison, are reasonably consistent with our results.} As a result, some of our outputs necessarily 
carry rather large uncertainty which stems directly from the large uncertainty associated with some of the input data. It is expected that experiments at the facilities such as DA$\Phi$NE, Frascati, will be providing data with such precision and quality that a more precise and thorough test of $CP$, $T$ and $CPT$ symmetries, and a meaningful test of the Bell-Steinberger relation itself as well, become possible.
%
%
%
%
\setcounter{section}{0}
\def\thesection{\Alph{section}}
\section{Appendix}
Denoting final states into which $|\ko\ket$ and $|\kob\ket$ decay generically as $|n\ket$, we consider the decay amplitudes
%
%
\begin{subequations}
\begin{eqnarray}
A_n &\equiv& \bra n|\Hw|\ko\ket \equiv |A_n|e^{i\psi_n} ~, \\ 
\Ab_{\nb} &\equiv& \bra \nb|\Hw|\kob\ket 
\equiv |\Ab_{\nb}|e^{i\psib_{\nb}}~,
\end{eqnarray}
\end{subequations}
where $|\nb\ket$ is, by definition, related to $|n\ket$ by
%
%
\begin{subequations}
\begin{eqnarray}
CP|n\ket = e^{i\al_n}|\nb\ket~, &\qquad& 
CPT|n\ket = e^{i\beta_n}|\nb\ket~, \\
CP|\nb\ket = e^{-i\al_n}|n\ket~, &\qquad& 
CPT|\nb\ket = e^{i\beta_n}|n\ket~.
\end{eqnarray}
\end{subequations}
With the aid of Eqs.(2.1) and (A.2a,b), one readily verifies that $CP$, $T$ and $CPT$ symmetries impose on the decay amplitudes $A_n$ and $\Ab_{\nb}$ such conditions as\footnote{It is understood that final state interactions may be neglected or have already been factored out.}
%
%
\begin{subequations}
\begin{eqnarray}
CP &\to& \Ab_{\nb} = A_n e^{-i(\al_K-\al_n)}~; \\
T  &\to& A_n^* = A_n e^{i(\beta_K-\beta_n-\al_K+\al_n)}~, \no \\
   &   & \Ab_{\nb}^* = \Ab_{\nb} e^{i(\beta_K-\beta_n+\al_K-\al_n)}~; \\
CPT &\to& \Ab_{\nb}^* = A_n e^{i(\beta_K-\beta_n)}~,
\end{eqnarray}
\end{subequations}
or
%
%
\begin{subequations}
\begin{eqnarray}
CP  &\to& |A_n| = |\Ab_{\nb}|~, \no \\
    &   & \psi_n - \psib_{\nb} + \al_n - \al_K = 0~; \\
T   &\to& 2\psi_n + \al_n - \beta_n - \al_K + \beta_K = 0~, \no \\
    &   & 2\psib_{\nb} - \al_n - \beta_n + \al_K + \beta_K = 0~; \\
CPT &\to& |A_n| = |\Ab_{\nb}|~, \no \\
    &   & \psi_n + \psib_{\nb} - \beta_n + \beta_K = 0~.
\end{eqnarray}
\end{subequations}
It is important to note that all these relations are invariant not only with respect to rephasing of $|\ko\ket$ and $|\kob\ket$, Eq.(4.14), but also with respect to rephasing of the final states,
%
%
\begin{equation}
|n\ket  \to |n\ket' = e^{-i\xi_n}|n\ket ~, \qquad |\nb\ket \to |\nb\ket' = e^{-i\xi_{\nb}}|\nb\ket ~,
\end{equation}
in spite that the phase parameters $\al_n$, $\beta_n$, $\psi_n$ and $\psib_{\nb}$ themselves are in general not invariant individually with respect to this rephasing. In fact, defining
%
%
\begin{subequations}
\begin{eqnarray}
CP|\ko\ket' = e^{i\al'_K}|\kob\ket' ~,&  &CPT|\ko\ket' = e^{i\beta'_K}|\kob\ket' ~, \\
CP|n\ket' = e^{i\al'_n}|\nb\ket' ~,&  &CPT|n\ket' = e^{i\beta'_n}|\nb\ket' ~, \\
'\bra n|\Hw|\ko\ket' = A'_n = |A_n|e^{i\psi'_n} ~,&  &'\bra \nb|\Hw|\kob\ket' =  \Ab'_{\nb} = |\Ab_{\nb}|e^{i\psib'_{\nb}} ~,
\end{eqnarray}
\end{subequations}
one finds
%
%
\begin{subequations}
\begin{eqnarray}
\al'_K = \al_K - 2\xi_K ~,&  &\beta'_K = \beta _K ~, \\ 
\al'_n = \al_n - \xi_n + \xi_{\nb} ~,&  &\beta'_n = \beta _n + \xi_n + \xi_{\nb} ~, \\
\psi'_n = \psi_n + \xi_n - \xi_K ~,&  &\psib'_{\nb} = \psib_{\nb} - \xi_{\nb} - \xi_K ~,
\end{eqnarray}
\end{subequations}
from which it follows that 
%
%
\begin{subequations}
\begin{eqnarray}
2\psi'_n + \al'_n - \beta'_n - \al'_K + \beta'_K &=& 
2\psi_n  + \al_n  - \beta_n  - \al_K  + \beta_K~, \\ 
2\psib'_{\nb} - \al'_n - \beta'_n + \al'_K + \beta'_K &=& 
2\psib_{\nb}  - \al_n  - \beta_n  + \al_K  + \beta_K~, \\
\psi'_n - \psib'_{\nb} + \al'_n - \al'_K &=& 
\psi_n  - \psib_{\nb}  + \al_n  - \al_K~, \\
\psi'_n + \psib'_{\nb} - \beta'_n + \beta'_K &=& 
\psi_n  + \psib_{\nb}  - \beta_n  + \beta_K~.
\end{eqnarray}
\end{subequations}

As $|n\ket$, we are interested in $|(2\pi)_I\ket$, $|\ell^ +\ket$ and $|\ell^-\ket$,\footnote{It is understood that $|\nb\ket \equiv |n\ket$, $\al_n \equiv 0$ and $\xi_{\nb} \equiv \xi_n$ for $|n\ket = |(2\pi)_I\ket$.} and, in Sect.4, we have parametrized the relevant 
amplitudes in a specific way, for which the conditions imposed by $CP$, $T$ and $CPT$ 
symmetries, Eqs.(A.4a,b,c), read
%
%
\begin{subequations}
\begin{eqnarray}
CP  &\to& \e_I = 0~, \qquad \theta_I = 0~, \no \\
    &   & \e_{\ell} = 0~, \qquad 2\theta_{\ell} + \al_{\ell} = 0~; \\
T   &\to& \theta_I = 0~, \qquad 2\phi_I - \beta_I + \beta_K = 0~, \no \\
    &   & 2\theta_{\ell} + \al_{\ell} = 0~, \qquad 2\phi_{\ell} - \beta_{\ell} + \beta_K = 0~; \\
CPT &\to& \e_I = 0~, \qquad 2\phi_I - \beta_I + \beta_K = 0~, \no \\
    &   & \e_{\ell} = 0~, \qquad 2\phi_{\ell} - \beta_{\ell} + \beta_K = 0~.
\end{eqnarray}
\end{subequations}
Note that all these constraints are independent of $\al_K$ and that the parameters $\e_I$, $\theta_I$ and $\e_{\ell}$ are each constrained to have a unique value, while the other 
parameters $\phi_I$, $\phi_{\ell}$ and $\theta_{\ell}$ are not. These three are just 
those phase parameters which are not invariant with respect to rephasing of the final states, Eq.(A.5), and accordingly may be transformed away by a rephasing. In fact, it is 
not difficult to see that freedom associated with choice of $\xi_I$, $\xi_{\ell+} + \xi_{\ell-}$ and $\xi_{\ell+} - \xi_{\ell-}$ allows one, without loss of generality, to take
%
%
\begin{equation}
\phi_I = 0~, \qquad \phi_{\ell} = 0~, \qquad \theta_{\ell} = 0~,
\end{equation}
respectively.

A couple of remarks are in order.

(1) Freedom associated with choice of $\xi_I$, $\xi_{\ell+} + \xi_{\ell-}$ and $\xi_{\ell+} - \xi_{\ell-}$ allows one alternatively to take 
%
%
\begin{equation}
\beta_I = \beta_K~, \qquad \beta_{\ell} = \beta_K~, \qquad \al_{\ell} = 0~,
\end{equation}
respectively. With this choice, one would, in addition to Eqs.(4.16a,b,c), have
%
%
\begin{subequations}
\begin{eqnarray}
CP  &\to& \theta_{\ell} = 0~; \\ 
T   &\to& \phi_I = 0~, \qquad \phi_{\ell} = 0~, \qquad \theta_{\ell} = 0~; \\
CPT &\to& \phi_I = 0~, \qquad \phi_{\ell} = 0~.
\end{eqnarray}
\end{subequations}
We prefer not to make this choice, since it would camouflage the fact that $\phi_I$, $\phi_{\ell}$ and $\theta_{\ell}$ are actually unmeasurable.

(2) In Refs.[8,9], parametrizing $A_I$, $\Ab_I$, $A_{\ell+}$ and $\Ab_{\ell-}$ as
%
%
\begin{subequations}
\begin{eqnarray}
A_I &=& F_I(1 + \e_I)e^{i\al_K/2}~, \\
\Ab_I &=& F_I(1 - \e_I)e^{-i\al_K/2}~, \\
A_{\ell+} &=& F_{\ell}(1 + \e_{\ell})e^{i\al_K/2}~, \\
\Ab_{\ell-} &=& F_{\ell}(1 - \e_{\ell})e^{-i\al_K/2}~,
\end{eqnarray}
\end{subequations}
with $F_I$, $\e_I$, $F_{\ell}$ and $\e_{\ell}$ all understood to be complex in general and with $\e_I$ and $\e_{\ell}$ presupposed to be small, we adopt a phase convention which gives
%
%
\begin{equation}
\im(F_I) = 0~, \qquad \im(F_{\ell}) = 0~, \qquad \im(\e_{\ell}) = 0~,
\end{equation}
so as to have
%
%
\begin{subequations}
\begin{eqnarray}
CP  &\to& \re(\e_I) = 0~, \qquad \im(\e_I) = 0~, \qquad \re(\e_{\ell}) = 0~; \\
T   &\to& \im(\e_I) = 0~; \\
CPT &\to& \re(\e_I) = 0~, \qquad \re(\e_{\ell}) = 0~.
\end{eqnarray}
\end{subequations}
Eq.(A.14) and Eqs.(A.15a,b,c) correspond to Eq.(4.15) and Eqs.(4.16a,b,c) respectively.\footnote{If one adopts an alternative phase convention which gives (A.11), one will have 
Eqs.(A.15a,b,c) and, in addition,
%
%
\begin{subequations}
\begin{eqnarray}
CP  &\to& \im(\e_{\ell}) = 0~; \\
T   &\to& \im(F_I) = 0~, \qquad \im(F_{\ell}) = 0~, \qquad \im(\e_{\ell}) = 0~; \\
CPT &\to& \im(F_I) = 0~, \qquad \im(F_{\ell}) = 0~.
\end{eqnarray}
\end{subequations}
}
We have preferred in this work to start with the parametrizations (4.10a,b) and (4.12a,b,c,d) so as to emphasize that all the phase parameters, $\phi_n$ and $\theta_n$ as well as $\al_K$, $\beta_K$, $\al_n$ and $\beta_n$, are completely arbitrary (i.e., not necessarily small) in general and it is rephasing-noninvariant phase parameters which may be transformed away by a rephasing.

(3) If, following Refs.[28,29], one parametrizes $A_{\ell+}$ and $\Ab_{\ell-}$ as
%
%
\begin{subequations}
\begin{eqnarray}
A_{\ell+} &=& f_{\ell}(1 - y_{\ell})~, \\
\Ab_{\ell-} &=& f_{\ell}^*(1 + y_{\ell}^*)~,
\end{eqnarray}
\end{subequations}
with both $f_{\ell}$ and $y_{\ell}$ understood to be complex in general, one would, corresponding to Eqs.(A.9a,b,c), have
%
%
\begin{subequations}
\begin{eqnarray}
CP  &\to& \re(y_{\ell}) = 0~, \qquad 2\im(f_{\ell})/\re(f_{\ell}) + \al_{\ell} - \al_K = 0~; \\
T   &\to& 2\im(f_{\ell})/\re(f_{\ell}) + \al_{\ell} - \al_K = 0~, \no \\
    &   & 2\im(y_{\ell}) - \beta_{\ell} + \beta_K = 0~; \\
CPT &\to& \re(y_{\ell}) = 0~, \qquad 2\im(y_{\ell}) - \beta_{\ell} + \beta_K = 0~.
\end{eqnarray}
\end{subequations}
This parametrization and our previous parametrization (A.13c,d), though distinct from each other in that the latter (the former) is invariant (not invariant) under rephasing of 
$|\ko\ket$ and $|\kob\ket$, Eq.(4.14), are similar to each other in that both are not 
invariant under rephasing of $|\ell^+\ket$ and $|\ell^-\ket$, Eq.(A.5). Specifically, both $\im(y_{\ell})$ and $\im(f_{\ell})$ (like $\im(F_{\ell})$ and $\im(\e_{\ell})$) are not 
invariant under the rephasing (A.5) and hence may be transformed away, leaving 
$\re(y_{\ell})$ (like $\re(\e_{\ell})$) as the only physical parameter characterizing $CP$ and $CPT$ violations.\footnote{The authors of Refs.[28,29] claimed that $CP$ and $CPT$ symmetries impose such 
conditions on $f_{\ell}$ and $y_{\ell}$ as
%
%
\begin{subequations}
\begin{eqnarray}
CP  &\to& \im(f_{\ell}) = 0~, \qquad \re(y_{\ell}) = 0~; \\
CPT &\to& \re(y_{\ell}) = 0~, \qquad \im(y_{\ell}) = 0~,
\end{eqnarray}
\end{subequations}
and argued that $\im(y_{\ell})$ does not appear, to first order, so that $y_{\ell}$ may be treated as real in their discussion. What we have argued here is, in contrast, that $CP~\to~\im(f_{\ell}) = 0$ and $CPT~\to~\im(y_{\ell}) = 0$ follow only if one adopts such phase convention as  $\al_{\ell} = \al_K$ and $\beta_{\ell} = \beta_K$ and that both $\im(f_{\ell})$ and $\im(y_{\ell})$ are to be regarded as parameters not measurable in principle.}

(4) The parameters $\e$ and $\de$ which characterize the $\ko-\kob$ mixing is often 
referred to as indirect symmetry-violating parameters while those parameters which 
characterize decay amplitudes are often referred to as direct symmetry-violating 
parameters. As is emphasized in [21], classification of symmetry-violating parameters into "direct" and "indirect" ones makes sense only when they are defined in such a way that 
they are invariant under rephasing of $|\ko\ket$ and $|\kob\ket$, Eq.(4.14).\footnote{In Ref.[7], we have further argumed that it is more consistent to refer to $\e _{\|}$ and $\de_{\perp}$, which are related exclusively to the mass matrix (see Eqs.(4.6a,d)), 
as indirect parameters and to all the other parameters including $\e _{\perp}$ and 
$\de_{\|}$ as direct parameters. Note also that direct parameters are not necessarily separable from indirect parameters.} This is the reason why we have been adhered to invariance under this rephasing.

(5) It is legitimate to parametrize $q_S/p_S$, $q_L/p_L$, $A_I$ and $\Ab_I$ in a way not invariant under the rephasing (4.14) and at the same time adopt some phase convention. For examples,
%
%
\begin{equation}
\left. 
	\begin{array}{c}
	\frac{\dps{q_S}}{\dps{p_S}} = \frac{\dps{1-\e-\de}}{\dps{1+\e+\de}}~, \vspace{2mm} \\
	\frac{\dps{q_L}}{\dps{p_L}} = \frac{\dps{1-\e+\de}}{\dps{1+\e-\de}}~,
	\end{array}
\right. 
\end{equation}
in place of Eqs.(4.1) and (4.2),
%
%
\begin{subequations}
\begin{eqnarray}
	A_I &=& F_I(1 + \e_I)e^{i(\phi_I + \theta_I)}~, \\
	\Ab_I &=& F_I(1 - \e_I)e^{i(\phi_I - \theta_I)}~,
\end{eqnarray}
\end{subequations}
in place of Eqs.(4.10a,b), and
%
%
\begin{subequations}
\begin{eqnarray}
	A_I &=& F_I(1 + \e_I)~, \\
	\Ab_I &=& F_I(1 - \e_I)~,
\end{eqnarray}
\end{subequations}
in place of Eqs.(A.13a,b). $\im(\e)$ and $\theta_I$ (or $\im(\e_I)$) then become 
noninvariant with respect to the rephasing (4.14). One may readily convince himself that 
freedom associated with choice of $\xi_K$ allows one, without loss of generality, to take either $\im(\e) = 0$ or, say, $\theta_0 = 0$ (or $\im(\e_0) = 0$). The latter is a phase 
convention corresponding to the one originally adopted by Wu and Yang [2], while the 
former is a phase convention once adopted by Wolfenstein [30]. Note that, although freedom associated with choice of $\xi_K$ also allows one, without loss of generality, to take $\al_K = 0$, which is a phase convention widely (sometimes implicitly, though) adopted in 
the literature, these three phase conventions are not compatible with one another, as has been emphasized before [21,24].
%
%

%
\end{document}